\documentclass[useAMS,usenatbib,onecolumn,referee]{mn2e}

\usepackage{dcolumn}
\usepackage{graphicx}
\usepackage{url}
\usepackage{color}
\usepackage{pdflscape}

\usepackage[table]{xcolor}

\newcommand{\cm}{cm$^{-1}$}

\newcommand{\X}{$X\,{}^3\Sigma^-$}
\newcommand{\astate}{$a\,{}^1\Delta$}

\newcommand{\ai}{\textit{ab initio}}

\newcommand{\eqref}[1]{(\ref{#1})}

\newcommand{\duo}{{\sc Duo}}
\newcommand{\Duo}{{\sc Duo}}

\title[ExoMol XXXI: The spectrum of PH]{ExoMol line list XXXIV: A Rovibrational Line List for Phosphinidene (PH) in its $X\,{}^3\Sigma^-$ and $a\,{}^1\Delta$ Electronic States}

\date{\today}
\author[Langleben  et al]{\large {Jonathan Langleben$^{1}$, Jonathan Tennyson$^{1}$\thanks{Email: j.tennyson@ucl.ac.uk}, Sergei N. Yurchenko$^{1}$ and Peter Bernath$^{2}$ }
\\
$^{1}$Department of Physics and Astronomy, University College London, London WC1E 6BT, UK \\
$^{2}$ Department of Chemistry and Biochemistry, Old Dominion University, 4541 Hampton Boulevard, Norfolk, VA 23529, USA
}

\date{Accepted XXXX. Received XXXX; in original form XXXX}

\pagerange{\pageref{firstpage}--\pageref{lastpage}} \pubyear{2018}

\begin{document}

\maketitle

\begin{abstract}

  A rovibronic line list for the ground ($X$~$^3\Sigma^-$) and first excited ($a$~$^1\Delta$) states of
  phosphinidene, $^{31}$PH, is computed.  The line list is designed for studies of exoplanetary and cool
  stellar atmospheres with temperatures up to 4000~K. A combination of
  empirical and \ai\ data are used to produce the line list:
 potential energy curves (PECs) are fitted using experimental transition frequencies;
these transitions are reproduced with a root mean square
  error of 0.01~cm$^{-1}$. The
  nuclear Schr\"{o}dinger is solved using these PECs plus Born-Oppenheimer
and spin splitting correction terms. Line
  intensities and Einstein~$A$ coefficients are computed using \ai\
  Dipole Moment Curves (DMC) $X$--$X$ and $a$--$a$. The resulting LaTY  line list, which
  contains 65055 transitions for 2528 rovibronic states up to 24500~\cm\ and $J=80$,
   is used to simulate spectra in emission and   absorption at a range of temperatures. The line list is
 made available in electronic form at the CDS and ExoMol databases.

\end{abstract}

\label{firstpage}

\section{Introduction}

The biochemistry of living organisms is heavily dependent on
phosphorus. It is important for the storage of genetic
information in the form of DNA and RNA (nucleic acids) and is a key
contributor to the structure of the cell membrane. The discovery of
phosphorus-bearing species in the astrophysical arena is thus thought to
be  of great
significance \citep{05Macia,17CuTiWa,18Zerkle}.
There have been numerous astrophysical attempts to
detect  diatomic phosphorus
molecules. To date, only PO \citep{07TeWoZi.PO}, PN
\citep{87Ziurys.PN} and CP \citep{90GuCePa.CP} have been identified in
the interstellar medium or circumstellar shells. The discovery of PH has so far eluded
astronomers \citep{80HoSnLo.PH,04HjBeBi.PH}.
However models of both the interstellar medium \citep{83ThAnPr.PH,90Millar.PH}
and (exo-) planetary atmospheres \citep{06ViLoFe.PH3} suggest
that there are environments where PH should be present in observable
quantities. The purpose of this paper is to provide a comprehensive
line list for PH to aid in its possible detection and modelling of
its spectrum. This line list supplements those of other phosphorus-bearing
molecules,
namely PN \citep{jt590}, PH$_3$ \citep{jt592}, PO and PS \citep{jt703},
produced as part of the ExoMol project \citep{jt528}.

The electronic ground state of PH, known as phosphinidene, is of
\X\ symmetry.
It  is a singly-bonded species so has a
lower dissociation energy, of $D_{\rm e}$(PH) $\approx 3.18(3)$ eV \citep{07Luo.book,18Rumble.book} (see detailed discussion below), compared
to the multiply bonded phosphorus species which have been detected in
space: $D_{\rm e}$(PN) $\approx{6.3}$~eV \citep{33CuHeHe.PN}, $D_{\rm e}$(PO) $\approx{5.47}$~eV
\citep{81RaReRa.PO} and $D_{\rm e}$(CP) $\approx{5.41}$~eV \citep{12ShXiSu.CP}.
The spectrum of PH has been well-studied in the laboratory
\citep{30Pearse.PH,39IsPexx.PH,74RoCoBa.PH,75DaRuTh.PH,78StLeMa.PH,78NgStLe.PH,84OhKaHi.PH,84AsDiSt.PH,84DrEn.PH,85GuKiLa.PH,87RaBe.PH,96BeSeSh.PH,96RaBexx.PH,97HuBrxx.PH,98KlKlWi.PH,99StLe.PH,03FiChWe.PH}
and its electronic structure has been the subject of a number of
computational studies \citep{81BrHiPe.PH,86SeRoWe.PH,92PaSu.PH,93GoSa.PH,02FiChMo.PH,12WaSuSh.PH,14GaGa.PH,13MuWo.PH}. As discussed below, a number of these works form key inputs to the present study.

\section{Method}

Our general methodology for constructing rotation-vibration
line lists is to use available experimental data to characterise the
underlying potential energy curve but to use
dipole moments computed {\it ab initio}; see \citet{jt511} and \citet{jt693}.
We follow this approach here. Since PH has a triplet electronic ground state,
it is necessary to supplement this approach with spin coupling terms
\citep{jt632}. All nuclear motion calculations are performed
with our general purpose, variational nuclear motion program for
diatomic molecules, \duo\ \citep{Duo}.
We give details of this procedure in the remainder of this section.

\subsection{Potential Energy Curve}

The \X\ and \astate\ PECs were represented using the
Extended Morse Oscillator (EMO) function as given by
\begin{equation}
\label{e:PEC}
V(r)=V_{\rm e}+ D_{\rm e} \Bigg[1-\exp\Bigg(-\sum^N_{k=0}\beta_k\xi_p^k(r-r_{\rm e})\Bigg)\Bigg]^2,
\end{equation}
where  $D_{\rm e}$ is the dissociation energy, $\xi_p$ is the $\check{S}$urkus variable given by
\begin{equation}
\xi_p=\Bigg(\frac{r^p-r^p_{\rm ref}}{r^p+r^p_{\rm ref}}\Bigg),
\end{equation}
$r_{\rm e}$ is the corresponding equilibrium bond length, $r_{\rm ref}$ defines the expansion centre for the $\xi_p$ variable (usually taken at $r_{\rm e}$) and the integer value $p$ influences how the function extrapolates beyond the data sensitive
region. This form allows for extra flexibility in the degree of the polynomial on the left or on the right sides of the
reference position, which is controlled by the parameters $N=$ $N_{l}$  and $N_{r}$, respectively.
 The empirical parameters $V_{\rm e}$ and $\beta$$_k$ are  derived through refinement to experimental data via a least-squares fit, while the dissociation energies $D_{\rm e}$ are constrained  to their experimental asymptotic energies $A_{\rm e}= V_{\rm e} + D_{\rm e}$ (see discussion below).

\begin{figure}
    \centering
  \includegraphics[width=0.85\textwidth]{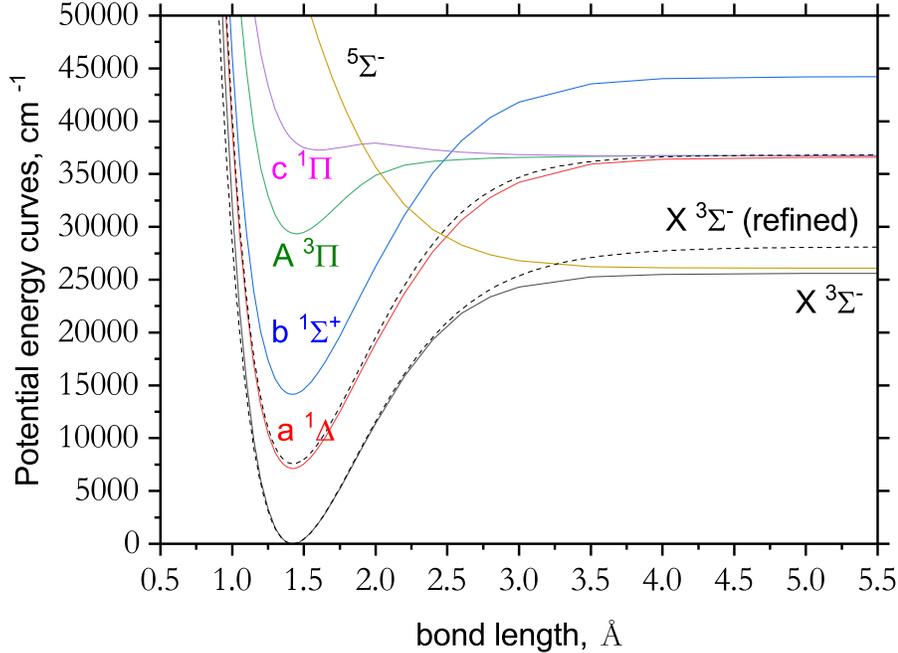}
    \label{f:PEC}
    \caption{\textit{Ab initio}  PECs of the lowest electronic states of PH from \protect\citet{14GaGa.PH} (solid lines)
    and the refined \X\ and \astate\ PECs  from this work (dashed).}
\end{figure}

A set of \ai\ MRCI+Q (Multi-Reference Configuration Interaction with
Davidson correction) curves for all low lying states of PH from
\citet{14GaGa.PH} is given in Fig.~\ref{f:PEC}. We used their \X\ and
\astate\ PECs as starting approximations for our model.  As there was
sufficient experimental data available for the
ground electronic states of PH \citep{87RaBe.PH,96RaBexx.PH,93GoSa.PH,98KlKlWi.PH}, the refined EMO PEC of 5th order was
essentially determined by empirical refinement rather than from the
\ai\ curve of \citet{14GaGa.PH}.  For \astate, only the rotational
lines for the $(0,0)$ band, $a$--$X$ electronic transition have been
characterized \citep{96BeSeSh.PH}. Therefore only $V_{\rm e}$ and
$r_{\rm e }$ parameters were refined, while all other parameters
defining the corresponding EMO PEC of 4th order were fixed to their
\ai\ values. The final set of potential parameters are listed in Table~\ref{t:PEC}.

\begin{table}
  \caption{Potential parameters defining empirical $X$ and $a$ PECs according to Eq. (\ref{e:PEC}). The units are \AA\ and \cm.}\label{t:PEC}
  \centering
  \begin{tabular}{lrr}
\hline
  Parameters& PEC ($X$)      & PEC($a$)        \\
  \hline
$ V_{\rm e}       $&$             0$&$           7569 $\\
$ r_{\rm e}       $&$      1.422179$&$       1.423911 $\\
$ D_{\rm e}       $&$         25700$&$          29431 $\\
$ r_{\rm ref}     $&$      1.422179$&$       1.423911 $\\
$ p               $&$             4$&$              4 $\\
$ N_{\rm L}       $&$             1$&$              1 $\\
$ N_{\rm R}       $&$             5$&$              4 $\\
$ B_0             $&$  1.7744781772$&$   1.6765776631 $\\
$ B_1             $&$  0.1557401848$&$  -0.0443775470 $\\
$ B_2             $&$  0.0119350618$&$   1.0241855379 $\\
$ B_3             $&$  0.4625698020$&$  -2.3099593616 $\\
$ B_4             $&$ -0.8284568722$&$   1.9917675261 $\\
$ B_5             $&$  0.9631507337$&$                $\\
\hline
\end{tabular}
\end{table}

Different couplings and corrections were modelled using the expansion
\begin{equation}
\label{e:bob}
F(r)= (1-\xi_p) \sum^{N}_{k=0}C_{k}\, z^{k}  + \xi_p\, C_{\infty v},
\end{equation}
where $z$ is either taken as the \v{S}urkus variable $z=\xi_p$  or using the damped-coordinate $z$ given by:
\begin{equation}\label{e:damp}
z = (r-r_{\rm ref})\, e^{-\gamma_2 (r-r_{\rm ref})^2-\gamma_4 (r - r_{\rm ref})^4},
\end{equation}
see also \citet{jt703} and \citet{jt711}. Here  $p$ (integer), $C_k$ and  $C_{\rm \infty v}$ are adjustable parameters. The expansion centre $r_{\rm ref}$ is typically chosen to be close to the equilibrium value of the ground electronic state.
To allow for rotational Born-Oppenheimer breakdown (BOB) effects \citep{Level} which become important
for high $J$, the vibrational kinetic energy operator for each state was extended by
\begin{equation}
\label{e:dist}
-\frac{\hbar^2}{2 \mu r^2} \to -\frac{\hbar^2}{2 \mu r^2} \left( 1 + g^{\rm BOB}(r)\right),
\end{equation}
where the unitless $X$ and $a$ BOB functions  were represented by the (unitless) polynomial $g^{\rm BOB}(r) = F(r)$.
The $X$-state BOB function required a 4th order expansion in terms of damped-$z$ in Eq.~\eqref{e:damp}, while for the $a$-stated only one $C_0$ constant from Eq.~\eqref{e:bob} was needed.
When fitting to the experimental frequencies, we had to include other $X$-state correction terms, such as spin-spin and spin-rotation, see \citet{Duo}, for which the same $z$-damped expression Eq.~(\ref{e:bob}) was used, of 3rd and 0th orders, respectively. All expansion parameters are given in Table~\ref{t:couplings} as well as in the supplementary material as part of the \Duo\ input file.

\begin{table}
  \caption{Expansion parameters defining spin-spin (SS), rotational Born-Oppenheimer breakdown (BOB) and spin-rotation (SR) functional forms according to Eqs.~(\ref{e:bob}) and (\ref{e:damp}).
  The bond length is in \AA, the SS and SR functions are in \cm, the BOB expansions are unitless. }\label{t:couplings}
  \centering
  \begin{tabular}{lrrclrr}
  \hline
   Eq. (\ref{e:damp})  &    SS ($X$)     &     BOB ($X$)    &   &  Eq. (\ref{e:bob})  & BOB ($a$)& SR ($X$)\\
   \hline
 $ r_{\rm ref}        $&$       1.42218 $&$         1.42218 $&  &$ r_{\rm ref}           $&$           1.42 $&$           1.42  $\\
 $ \gamma_2         $&$           0.8 $&$             0.8 $&  &$ p                   $&$              4 $&$              4  $\\
 $ \gamma_4         $&$             0 $&$               0 $&  &$ C_0                 $&$   0.0047811183 $&$  -0.0797377811  $\\
 $ p                $&$             2 $&$               2 $&  &$ C_{\infty}            $&$              0 $&$              0  $\\
 $ C_0              $&$  1.4728278045 $&$   -0.0001063699 $&  &$                     $&$                $&$                 $\\
 $ C_1              $&$  1.0957652053 $&$    0.0089251295 $&  &$                     $&$                $&$                 $\\
 $ C_2              $&$  0.4850903021 $&$   -0.3045117774 $&  &$                     $&$                $&$                 $\\
 $ C_3              $&$             0 $&$    0.3645474486 $&  &$                     $&$                $&$                 $\\
 $ C_4              $&$             0 $&$   -0.3901575147 $&  &$                     $&$                $&$                 $\\
 $ C_{\infty}         $&$             0 $&$               0 $&  &$                     $&$                $&$                 $\\
\hline
\end{tabular}
\end{table}

Experimental data was taken primarily from four sources.  To determine
the $X$-state curves, submillimeter-wave measurements of the  ($N'=0,1$)
rotational spectrum by \citet{93GoSa.PH} and the ($N'=1,2$) spectrum
by \citet{98KlKlWi.PH} were combined with frequencies from an infrared
Fourier transform spectrometer vibration-rotation spectrum of the
ground state of PH by \citet{87RaBe.PH}. The infrared study observed
five vibrational bands (1--0, 2--1, 3--2, 4--3 and 5--4) up to a
maximum rotation state of $J=21$. The $X$-state data set comprised 381
lines split between six fine-structure resolved branches: $R_1$,
$R_2$, $R_3$, $P_1$, $P_2$ and $P_3$ which characterize the triplet
pattern arising from the splitting of the lines from the electronic
spin angular momentum along the internuclear axis (i.e. different
spin-projections $\Sigma= -1, 0, 1$; note that the $X$ state is Hund's
case (b) so these projections on the internuclear axis are not good
quantum numbers). These data allowed the lowest 6 vibrational states
($v=0\ldots 5$) to be characterised. The hyperfine structure of the
submillimeter-wave frequencies \citep{93GoSa.PH,98KlKlWi.PH} were
averaged.  For the $a$-state, 64 $a$--$X$ IR transition frequencies
(0--0 band) from \cite{96BeSeSh.PH} were used.

According to \citet{18Rumble.book} (with the original reference to \citet{07Luo.book}), the dissociation value for PH at $T=298$~K is  297.0$\pm$2.1~kJ/mole or 293.2 kJ/mole at 0~K (24~516$\pm$175~\cm\ or 3.04(2) eV).   Using a zero point energy value of 1170~\cm\ from our calculations (ZPE = $D_{\rm e} - D_{\rm 0}$), this corresponds to $D_{\rm e} = $ 25~687$\pm$175~\cm\ (3.19~eV).
The value $A_{\rm e} (X) = D_{\rm e}(X) = 25~700$~\cm\ was adopted as the dissociation energy of the $X$ state in our calculations.
Throughout the refinement phase, the equilibrium distance $r_{\rm e}$
was kept fixed at $r_{\rm e}$ = 1.4221~\AA, as spectroscopically determined by \citet{87RaBe.PH}.

The dissociation channel for the $a$ states can be estimated using the phosphorus atom excitation energy ($^2D_{3/2}$):
 $A_{\rm e}(a) = D_{\rm e}(X) + 11 361.02 $  $\approx$   37000~\cm \citep{NIST}, which was used to constrain the dissociation energy of \astate\ while varying the corresponding value of the origin $V_{\rm e}$.

\begin{figure}
    \centering
   \includegraphics[width=0.47\textwidth]{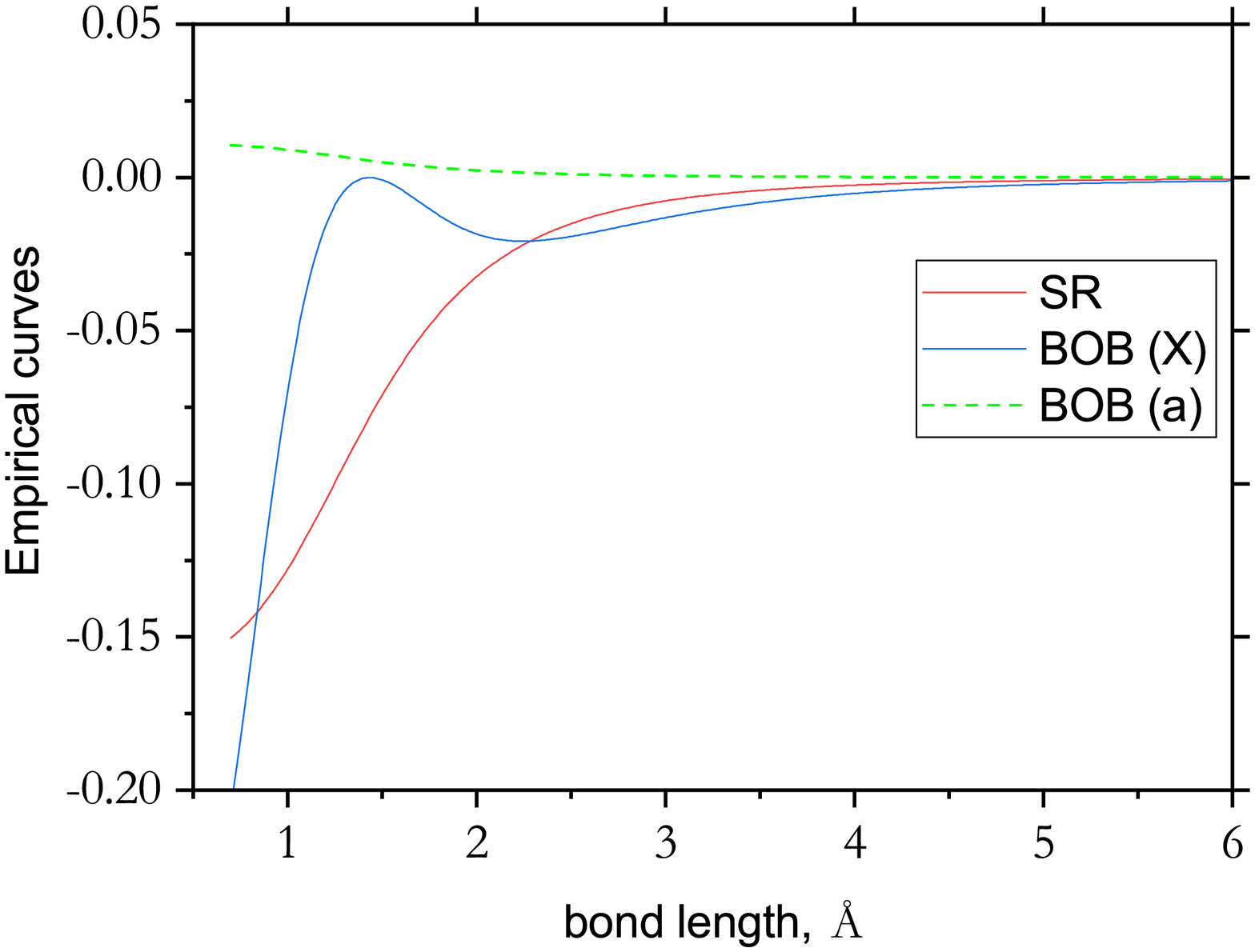}
   \includegraphics[width=0.47\textwidth]{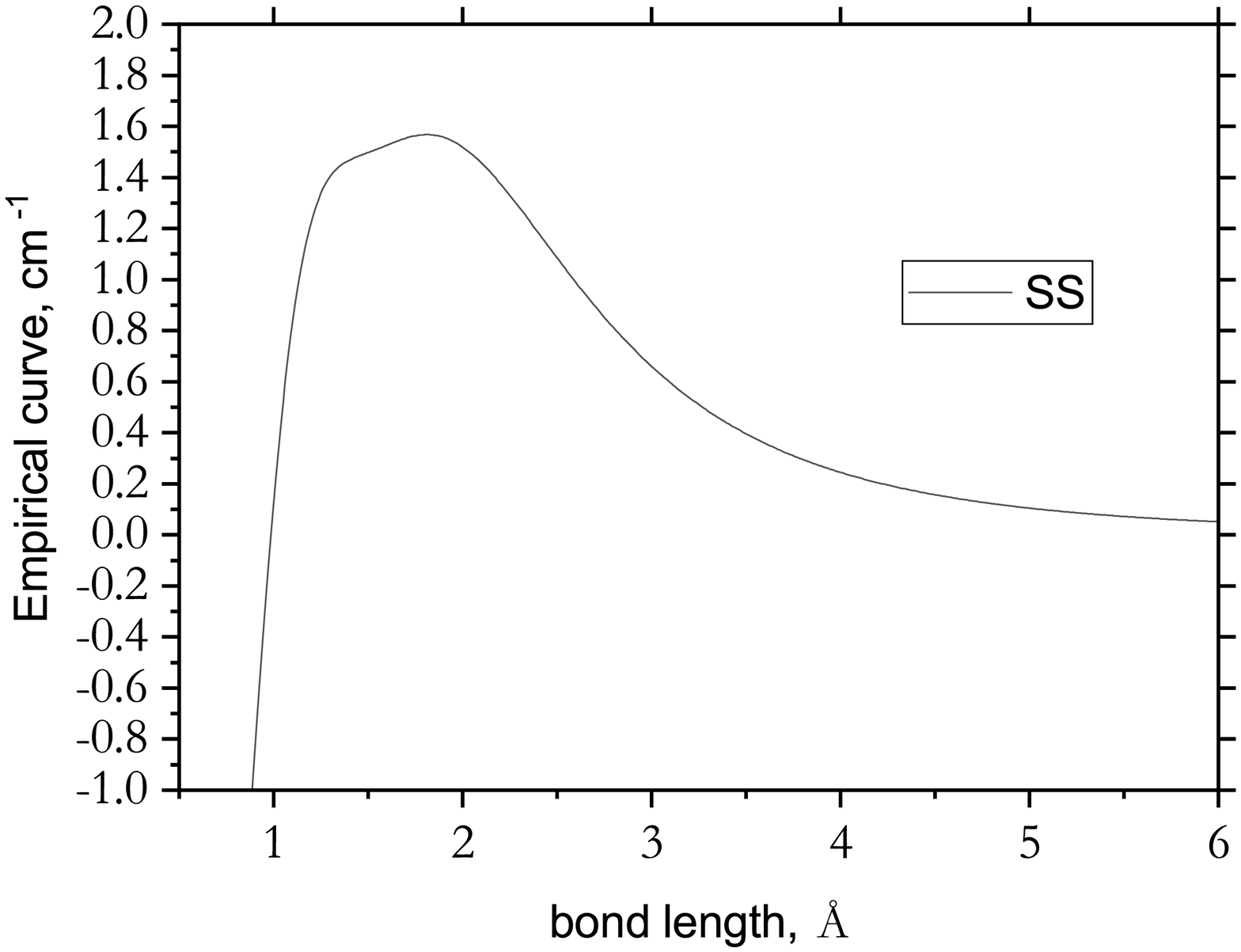}
    \caption{Empirical curves of the \X\ state of PH from in this work: spin-spin (\cm), spin-rotation (\cm) and Born-Oppenheimer breakdown (unitless).  }
    \label{f:SS-SR-BOB}
\end{figure}

The PECs as well as other empirical curves fitted directly to experimental-measured frequencies ($J\le 21$) achieved a root-mean-square error of 0.01 cm$^{-1}$. The final refined PECs are shown in Fig.~\ref{f:PEC} and empirical curves are shown in Fig.~\ref{f:SS-SR-BOB}. Some of the residuals are illustrated in Tables~\ref{t:vib}--\ref{t:X-A}.

\begin{table}\centering
\caption{A comparison of experimentally derived vibrational energy levels
 (in cm$^{-1}$) for the \X\ state of $^{31}$PH  from  \protect\citet{96RaBexx.PH} (Obs.)
with those calculated using \Duo\ (Calc.).}\label{t:vib}
\begin{tabular}{crrr}
\hline
          $v$          &   Obs.    &      Calc.    &        Obs.-Calc.     \\
\hline
       1   &    2276.20901(51)      &          2276.2061   &     0.00291   \\
       2   &    4465.02033(74)      &          4465.0148   &     0.00553   \\
       3   &    6566.15898(88)      &          6566.1561   &     0.00288   \\
       4   &    8578.9443(11)       &          8578.9506   &     -0.0063   \\
       5   &    10502.1949(13)      &         10502.2006   &     -0.0057   \\
\hline
\end{tabular}
\end{table}

\begin{table}\centering
\caption{A comparison of experimental sub-millimeter frequencies (in cm$^{-1}$) for the \X\ state of $^{31}$PH from  \protect\citet{93GoSa.PH} and \protect\citet{98KlKlWi.PH} (Obs.)
with those calculated using \Duo\ (Calc.). }\label{t:milli}

\begin{tabular}{rccrrccrrrr}
\hline
          $J'$ & $\pm'$ & $e/f'$ & $N'$ & $J''$ & $\pm''$ & $e/f''$  & $N''$   &   Obs.    &      Calc.    &        Obs.-Calc.     \\
\hline
       0 &     +     &     e     &         1 &         1 &  -        &  e        &         0 &   14.1191 &   14.1185 &    0.0006 \\
       1 &     +     &     f     &         1 &         1 &  -        &  e        &         0 &   18.4634 &   18.4623 &    0.0011 \\
       2 &     +     &     e     &         1 &         1 &  -        &  e        &         0 &   16.4813 &   16.4811 &    0.0002 \\
       1 &     -     &     e     &         2 &         1 &  +        &  f        &         1 &   30.9308 &   30.9299 &    0.0009 \\
       1 &     -     &     e     &         2 &         0 &  +        &  e        &         1 &   35.2757 &   35.2738 &    0.0019 \\
       2 &     -     &     f     &         2 &         1 &  +        &  f        &         1 &   33.6363 &   33.6336 &    0.0027 \\
       2 &     -     &     f     &         2 &         2 &  +        &  e        &         1 &   35.6146 &   35.6148 &   -0.0002 \\
       3 &     -     &     e     &         2 &         2 &  +        &  e        &         1 &   33.4474 &   33.4445 &    0.0029 \\
\hline
\end{tabular}
\end{table}

\begin{table}\centering
\caption{A comparison of experimental IR frequencies (in cm$^{-1}$)
for the \X\ state of $^{31}$PH from \protect\citet{87RaBe.PH} (Obs.)
with those calculated using \Duo\ (Calc.) for a selection of transitions.}
\label{t:X-X}
\begin{tabular}{rccrrrccrrrrr}
\hline
          $J'$ & $\pm'$ &$e/f'$ & $v'$ & $N'$ & $J''$  & $\pm''$ &$e/f''$  & $v''$ & $N''$    &   Obs.    &      Calc.    &        Obs.-Calc.     \\
\hline
   0 &   +   &   e   &  1   &  1  &  1   &   -   &  e    &        0 &        2 &  2240.4245 &   2240.4228  &    0.0017   \\
   0 &   +   &   e   &  2   &  1  &  1   &   -   &  e    &        1 &        2 &  2154.0349 &   2154.0248  &    0.0101   \\
   0 &   +   &   e   &  3   &  1  &  1   &   -   &  e    &        2 &        2 &  2067.3792 &   2067.3730  &    0.0062   \\
   1 &   -   &   e   &  1   &  2  &  0   &   +   &  e    &        0 &        1 &  2309.9615 &   2309.9625  &   -0.0010   \\
   1 &   -   &   e   &  1   &  0  &  2   &   +   &  e    &        0 &        1 &  2259.7221 &   2259.7249  &   -0.0028   \\
   1 &   +   &   f   &  1   &  1  &  2   &   -   &  f    &        0 &        1 &  2242.0659 &   2242.0688  &   -0.0029   \\
   1 &   -   &   e   &  1   &  2  &  2   &   +   &  e    &        0 &        3 &  2223.9221 &   2223.9240  &   -0.0019   \\
   1 &   -   &   e   &  2   &  0  &  2   &   +   &  e    &        1 &        1 &  2172.8223 &   2172.8137  &    0.0086   \\
   1 &   +   &   f   &  2   &  1  &  2   &   -   &  f    &        1 &        1 &  2155.6753 &   2155.6634  &    0.0119   \\
   1 &   -   &   e   &  2   &  2  &  2   &   +   &  e    &        1 &        3 &  2138.0416 &   2138.0324  &    0.0092   \\
   1 &   -   &   e   &  3   &  0  &  2   &   +   &  e    &        2 &        1 &  2085.6489 &   2085.6481  &    0.0008   \\
   1 &   +   &   f   &  3   &  1  &  2   &   -   &  f    &        2 &        1 &  2069.0122 &   2069.0003  &    0.0119   \\
   1 &   -   &   e   &  3   &  2  &  2   &   +   &  e    &        2 &        3 &  2051.8808 &   2051.8782  &    0.0026   \\
\ldots \\
   5 &   +   &   f   &  3   &  5  &  6   &   -   &  f    &        2 &        5 &  1999.0172 &   1999.0052  &    0.0120   \\
   5 &   -   &   e   &  3   &  6  &  6   &   +   &  e    &        2 &        7 &  1980.2734 &   1980.2631  &    0.0103   \\
   5 &   -   &   e   &  4   &  4  &  4   &   +   &  e    &        3 &        3 &  2068.7136 &   2068.7384  &   -0.0248   \\
   5 &   +   &   f   &  4   &  5  &  4   &   -   &  f    &        3 &        3 &  2081.4480 &   2081.4679  &   -0.0199   \\
   5 &   -   &   e   &  4   &  6  &  4   &   +   &  e    &        3 &        5 &  2093.6318 &   2093.6574  &   -0.0256   \\
   5 &   -   &   e   &  4   &  4  &  6   &   +   &  e    &        3 &        5 &  1931.4560 &   1931.4863  &   -0.0303   \\
   5 &   +   &   f   &  4   &  5  &  6   &   -   &  f    &        3 &        5 &  1913.6358 &   1913.6587  &   -0.0229   \\
   5 &   -   &   e   &  4   &  6  &  6   &   +   &  e    &        3 &        7 &  1895.3895 &   1895.4050  &   -0.0155   \\
   5 &   -   &   e   &  5   &  4  &  4   &   +   &  e    &        4 &        3 &  1977.0756 &   1977.0671  &    0.0085   \\
   5 &   +   &   f   &  5   &  5  &  4   &   -   &  f    &        4 &        3 &  1989.2557 &   1989.2403  &    0.0154   \\
   5 &   -   &   e   &  5   &  6  &  4   &   +   &  e    &        4 &        5 &  2000.8809 &   2000.8782  &    0.0027   \\
   6 &   +   &   e   &  1   &  5  &  5   &   -   &  e    &        0 &        4 &  2352.4659 &   2352.4577  &    0.0082   \\
   6 &   -   &   f   &  1   &  6  &  5   &   +   &  f    &        0 &        4 &  2366.1885 &   2366.1869  &    0.0016   \\
\ldots \\
  10 &   +   &   e   &  2   & 11  &  11  &   -   &  e    &        1 &       12 &  1962.6171 &   1962.6219  &   -0.0048   \\
  10 &   +   &   e   &  3   &  9  &  9   &   -   &  e    &        2 &        8 &  2219.3963 &   2219.3746  &    0.0217   \\
  10 &   -   &   f   &  3   & 10  &  9   &   +   &  f    &        2 &        8 &  2229.7211 &   2229.7033  &    0.0178   \\
  10 &   +   &   e   &  3   & 11  &  9   &   -   &  e    &        2 &       10 &  2239.4441 &   2239.4372  &    0.0069   \\
  10 &   +   &   e   &  3   &  9  &  11  &   -   &  e    &        2 &       10 &  1922.0451 &   1922.0214  &    0.0237   \\
  10 &   -   &   f   &  3   & 10  &  11  &   +   &  f    &        2 &       10 &  1901.6330 &   1901.6003  &    0.0327   \\
  10 &   +   &   e   &  3   & 11  &  11  &   -   &  e    &        2 &       12 &  1880.8366 &   1880.8019  &    0.0347   \\
  10 &   +   &   e   &  4   &  9  &  9   &   -   &  e    &        3 &        8 &  2126.2936 &   2126.3060  &   -0.0124   \\
  10 &   -   &   f   &  4   & 10  &  9   &   +   &  f    &        3 &        8 &  2136.0604 &   2136.0738  &   -0.0134   \\
  10 &   +   &   e   &  4   & 11  &  9   &   -   &  e    &        3 &       10 &  2145.2285 &   2145.2524  &   -0.0239   \\
  10 &   +   &   e   &  5   &  9  &  9   &   -   &  e    &        4 &        8 &  2031.8406 &   2031.8191  &    0.0215   \\
  10 &   -   &   f   &  5   & 10  &  9   &   +   &  f    &        4 &        8 &  2041.0076 &   2040.9884  &    0.0192   \\
\ldots\\
  16 &   +   &   e   &  3   & 15  &  15  &   -   &  e    &        2 &       14 &  2271.8427 &   2271.8199  &    0.0228   \\
  16 &   -   &   f   &  3   & 16  &  15  &   +   &  f    &        2 &       14 &  2278.3547 &   2278.3273  &    0.0274   \\
  16 &   +   &   e   &  3   & 17  &  15  &   -   &  e    &        2 &       16 &  2284.1877 &   2284.1849  &    0.0028   \\
  16 &   +   &   e   &  4   & 15  &  15  &   -   &  e    &        3 &       14 &  2175.3368 &   2175.3445  &   -0.0077   \\
  17 &   -   &   e   &  1   & 16  &  16  &   +   &  e    &        0 &       15 &  2469.8998 &   2469.8935  &    0.0063   \\
  17 &   +   &   f   &  1   & 17  &  16  &   -   &  f    &        0 &       15 &  2476.8083 &   2476.8153  &   -0.0070   \\
  17 &   -   &   e   &  1   & 18  &  16  &   +   &  e    &        0 &       17 &  2483.0468 &   2483.0661  &   -0.0193   \\
  17 &   -   &   e   &  1   & 16  &  18  &   +   &  e    &        0 &       17 &  1930.5765 &   1930.5700  &    0.0065   \\
  17 &   +   &   f   &  1   & 17  &  18  &   -   &  f    &        0 &       17 &  1906.7108 &   1906.7114  &   -0.0006   \\
  17 &   -   &   e   &  1   & 18  &  18  &   +   &  e    &        0 &       19 &  1882.5218 &   1882.5467  &   -0.0249   \\
  17 &   -   &   e   &  2   & 16  &  16  &   +   &  e    &        1 &       15 &  2374.4089 &   2374.4329  &   -0.0240   \\
  17 &   +   &   f   &  2   & 17  &  16  &   -   &  f    &        1 &       15 &  2380.8107 &   2380.8373  &   -0.0266   \\
  17 &   -   &   e   &  2   & 18  &  16  &   +   &  e    &        1 &       17 &  2386.5164 &   2386.5774  &   -0.0610   \\
  17 &   -   &   e   &  3   & 16  &  16  &   +   &  e    &        2 &       15 &  2278.3547 &   2278.3206  &    0.0341   \\
  17 &   +   &   f   &  3   & 17  &  16  &   -   &  f    &        2 &       15 &  2284.1877 &   2284.1579  &    0.0298   \\
  17 &   -   &   e   &  3   & 18  &  16  &   +   &  e    &        2 &       17 &  2289.3440 &   2289.3366  &    0.0074   \\
  18 &   +   &   e   &  1   & 17  &  17  &   -   &  e    &        0 &       16 &  2476.8083 &   2476.7978  &    0.0105   \\
  18 &   -   &   f   &  1   & 18  &  17  &   +   &  f    &        0 &       16 &  2483.0468 &   2483.0481  &   -0.0013   \\
 \hline
\end{tabular}
\end{table}

\begin{table}\centering
\label{t:X-A}
\caption{A comparison of experimental \astate\ -- \X\ (forbidden) frequencies (in cm$^{-1}$) from \protect\citet{96BeSeSh.PH} (Obs.)
with those calculated using \Duo\ (Calc.). }
\begin{tabular}{rccrrrccrrrrr}
\hline
          $J'$ & $\pm'$ &$e/f'$ & $v'$ & $N'$ & $J''$  & $\pm''$ &$e/f''$  & $v''$ & $N''$    &   Obs.    &      Calc.    &        Obs.-Calc.     \\
\hline
   2 &   +   &   e   &  0   &  2  &  3   &   -   &  e    &        0 &        4 &  7406.0737 &   7406.0727  &    0.0010\\
   3 &   -   &   e   &  0   &  3  &  4   &   +   &  e    &        0 &        5 &  7372.6299 &   7372.6427  &   -0.0128\\
   4 &   +   &   e   &  0   &  4  &  5   &   -   &  e    &        0 &        6 &  7339.3906 &   7339.3921  &   -0.0015\\
   5 &   -   &   e   &  0   &  5  &  6   &   +   &  e    &        0 &        7 &  7306.3367 &   7306.3324  &    0.0043\\
   6 &   +   &   e   &  0   &  6  &  7   &   -   &  e    &        0 &        8 &  7273.4977 &   7273.4805  &    0.0172\\
   7 &   -   &   e   &  0   &  7  &  8   &   +   &  e    &        0 &        9 &  7240.8720 &   7240.8559  &    0.0161\\
   8 &   +   &   e   &  0   &  8  &  9   &   -   &  e    &        0 &       10 &  7208.5009 &   7208.4794  &    0.0215\\
   9 &   -   &   e   &  0   &  9  &  10  &   +   &  e    &        0 &       11 &  7176.3990 &   7176.3726  &    0.0264\\
   2 &   -   &   f   &  0   &  2  &  3   &   +   &  f    &        0 &        2 &  7471.0723 &   7471.0815  &   -0.0092\\
   3 &   +   &   f   &  0   &  3  &  4   &   -   &  f    &        0 &        3 &  7454.5009 &   7454.5073  &   -0.0064\\
   4 &   -   &   f   &  0   &  4  &  5   &   +   &  f    &        0 &        4 &  7438.0317 &   7438.0375  &   -0.0058\\
   5 &   +   &   f   &  0   &  5  &  6   &   -   &  f    &        0 &        5 &  7421.6792 &   7421.6832  &   -0.0040\\
   6 &   -   &   f   &  0   &  6  &  7   &   +   &  f    &        0 &        6 &  7405.4577 &   7405.4557  &    0.0020\\
   7 &   +   &   f   &  0   &  7  &  8   &   -   &  f    &        0 &        7 &  7389.3738 &   7389.3667  &    0.0071\\
   8 &   -   &   f   &  0   &  8  &  9   &   +   &  f    &        0 &        8 &  7373.4400 &   7373.4278  &    0.0122\\
   9 &   +   &   f   &  0   &  9  &  10  &   -   &  f    &        0 &        9 &  7357.6640 &   7357.6514  &    0.0126\\
   2 &   +   &   e   &  0   &  2  &  2   &   -   &  f    &        0 &        1 &  7521.4973 &   7521.5055  &   -0.0082\\
   3 &   -   &   e   &  0   &  3  &  3   &   +   &  f    &        0 &        2 &  7521.6843 &   7521.6904  &   -0.0061\\
   4 &   +   &   e   &  0   &  4  &  4   &   -   &  f    &        0 &        3 &  7521.9341 &   7521.9375  &   -0.0034\\
   5 &   -   &   e   &  0   &  5  &  5   &   +   &  f    &        0 &        4 &  7522.2483 &   7522.2477  &    0.0006\\
   6 &   +   &   e   &  0   &  6  &  6   &   -   &  f    &        0 &        5 &  7522.6249 &   7522.6217  &    0.0032\\
   7 &   -   &   e   &  0   &  7  &  7   &   +   &  f    &        0 &        6 &  7523.0669 &   7523.0608  &    0.0061\\
   8 &   +   &   e   &  0   &  8  &  8   &   -   &  f    &        0 &        7 &  7523.5775 &   7523.5663  &    0.0112\\
   9 &   -   &   e   &  0   &  9  &  9   &   +   &  f    &        0 &        8 &  7524.1112 &   7524.1400  &   -0.0288\\
  10 &   +   &   e   &  0   & 10  &  10  &   -   &  f    &        0 &        9 &  7524.7853 &   7524.7841  &    0.0012\\
   2 &   -   &   f   &  0   &  2  &  1   &   +   &  f    &        0 &        0 &  7555.1330 &   7555.1391  &   -0.0061\\
   3 &   +   &   f   &  0   &  3  &  2   &   -   &  f    &        0 &        1 &  7572.1114 &   7572.1144  &   -0.0030\\
   4 &   -   &   f   &  0   &  4  &  3   &   +   &  f    &        0 &        2 &  7589.1204 &   7589.1206  &   -0.0002\\
   5 &   +   &   f   &  0   &  5  &  4   &   -   &  f    &        0 &        3 &  7606.1501 &   7606.1477  &    0.0024\\
   6 &   -   &   f   &  0   &  6  &  5   &   +   &  f    &        0 &        4 &  7623.1921 &   7623.1862  &    0.0059\\
   7 &   +   &   f   &  0   &  7  &  6   &   -   &  f    &        0 &        5 &  7640.2333 &   7640.2267  &    0.0066\\
   8 &   -   &   f   &  0   &  8  &  7   &   +   &  f    &        0 &        6 &  7657.2646 &   7657.2604  &    0.0042\\
   9 &   +   &   f   &  0   &  9  &  8   &   -   &  f    &        0 &        7 &  7674.2851 &   7674.2784  &    0.0067\\
  10 &   -   &   f   &  0   & 10  &  9   &   +   &  f    &        0 &        8 &  7691.2735 &   7691.2726  &    0.0009\\
  11 &   +   &   f   &  0   & 11  &  10  &   -   &  f    &        0 &        9 &  7708.2094 &   7708.2353  &   -0.0259\\
  12 &   -   &   f   &  0   & 12  &  11  &   +   &  f    &        0 &       10 &  7725.1395 &   7725.1594  &   -0.0199\\
   2 &   +   &   e   &  0   &  2  &  1   &   -   &  e    &        0 &        0 &  7573.5937 &   7573.6014  &   -0.0077\\
   3 &   -   &   e   &  0   &  3  &  2   &   +   &  e    &        0 &        1 &  7607.7241 &   7607.7292  &   -0.0051\\
   4 &   +   &   e   &  0   &  4  &  3   &   -   &  e    &        0 &        2 &  7641.7120 &   7641.7149  &   -0.0029\\
   5 &   -   &   e   &  0   &  5  &  4   &   +   &  e    &        0 &        3 &  7675.6392 &   7675.6403  &   -0.0011\\
   6 &   +   &   e   &  0   &  6  &  5   &   -   &  e    &        0 &        4 &  7709.5131 &   7709.5125  &    0.0006\\
   7 &   -   &   e   &  0   &  7  &  6   &   +   &  e    &        0 &        5 &  7743.3243 &   7743.3220  &    0.0023\\
   8 &   +   &   e   &  0   &  8  &  7   &   -   &  e    &        0 &        6 &  7777.0560 &   7777.0542  &    0.0018\\
   9 &   -   &   e   &  0   &  9  &  8   &   +   &  e    &        0 &        7 &  7810.6816 &   7810.6924  &   -0.0108\\
  10 &   +   &   e   &  0   & 10  &  9   &   -   &  e    &        0 &        8 &  7844.1942 &   7844.2194  &   -0.0252\\
  10 &   -   &   f   &  0   & 10  &  11  &   +   &  f    &        0 &       10 &  7342.0698 &   7342.0500  &    0.0198\\
\hline
\end{tabular}
\end{table}

\subsection{\textit{Ab initio} Dipole Moment Curves}

\begin{figure}
    \centering
  \includegraphics[width=0.9\textwidth]{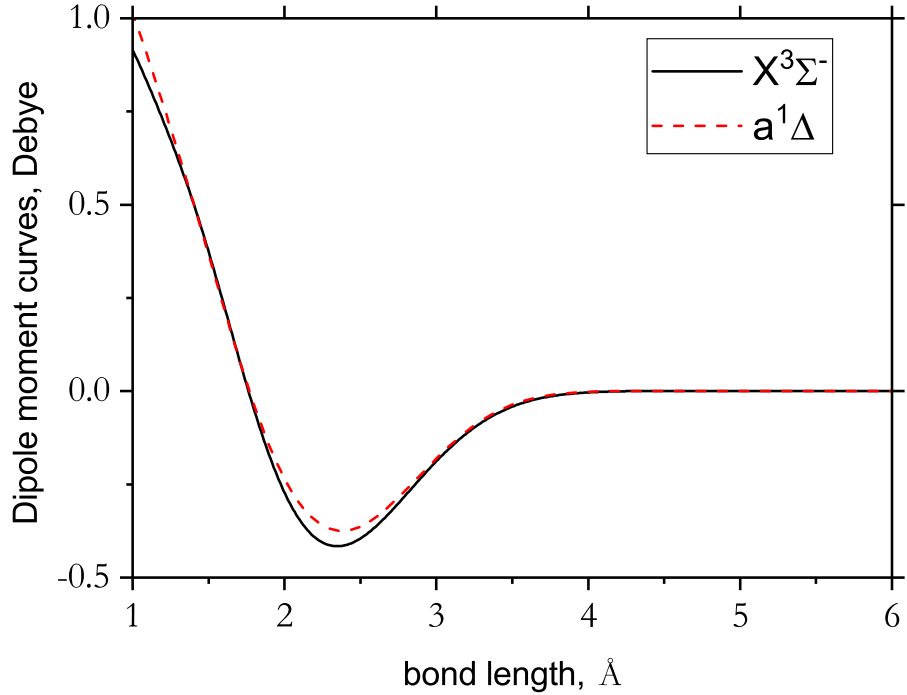}
    \label{f:DMC}
    \caption{{\it Ab initio} PH MRCI + Q / AV5Z Dipole Moment Curve (DMC)
used in this work: the \X\ state DMC is
    from \protect\citet{14GaGa.PH} and \astate\ DMC was computed
as part of this study.}
\end{figure}

PH has a permanent dipole and can be represented by a dipole moment
curve (DMC) which shows the variation of the dipole with internuclear
separation. \citet{14GaGa.PH} computed an \ai\ DMC for the
\X\ state of PH using an aug-ccpV5Z basis set at the MRCI+Q
level of theory. We used the same method to compute the DMC for the \astate\ state with the MOLPRO program \citep{MOLPRO}. These DMCs are illustrated in Fig. \ref{f:DMC}. As
expected for a neutral species, the dipole moments tend to zero at large bond lengths.

The \X\-state \ai\ dipole moment has an equilibrium value of 0.4771~D. This compares reasonably with other theoretical study equilibrium values. \citet{13MuWo.PH} calculated dipole moments at a number of theoretical levels, using a variety of basis sets. An AV5Z
basis set at the coupled cluster (CCSD(T)) level of theory produced a
dipole moment of 0.4410~D. Other studies at the
configuration-interaction (CI) level calculated the dipole moment as
0.431~D \citep{86SeRoWe.PH} and 0.432~D \citep{92PaSu.PH}.
\citet{75MeRo.PH} using the coupled electron pair-approach (CEPA)
produced a dipole moment of 0.481~D, which is closest to the
value of \citet{14GaGa.PH} used for our line list calculations.

The DMCs were represented analytically using the damped-$z$ expansion given by Eq.~\eqref{e:bob}.
This was done in order to reduce the numerical noise in the calculated intensities for
high overtones; see recommendations by \citet{16MeMeSt}.
These DMCs  and the empirically defined curves constitute our spectroscopic model, which was then used with \duo\ to produce a PH line list. The corresponding expansion parameters are listed in Table~\ref{t:DMC}. They can also be found in the supplementary data in the form of the \duo\ input files together with the corresponding grid representations.

\begin{table}
  \caption{Expansion parameters defining $X$ and $a$ DMCs according to Eq.~(\ref{e:damp}). The units are \AA\ and Debye.}\label{t:DMC}
  \centering
  \begin{tabular}{lrr}
\hline
Parameters  &      DMC($X$)     &     DMC($a$)   \\
  \hline
$ r_{\rm ref}       $&$          1.4222 $&$         1.4222$\\
$ \gamma_2          $&$    0.1695549437 $&$   0.1233507271$\\
$ \gamma_4          $&$    0.0116033136 $&$   0.0210533244$\\
$ p                 $&$               6 $&$              6$\\
$ D_0               $&$    0.4766434841 $&$   0.4747489601$\\
$ D_1               $&$   -0.2603505526 $&$  -0.3886599811$\\
$ D_2               $&$   -1.6119565470 $&$  -1.3959503273$\\
$ D_3               $&$   -4.4208750252 $&$  -3.5580088368$\\
$ D_4               $&$   -3.6843015061 $&$  -2.7131260560$\\
$ D_{\infty}          $&$               0 $&$              0$\\
\hline
\end{tabular}
\end{table}

\section{Results}

\subsection{Line list}

The spectroscopic model described in the previous sections was used to
generate the line list for the ground ($X$) and first excited ($a$)
electronic states of $^{31}$PH. The LaTY line list was computed using
the empirical PECs and correction curves and \ai\ DMCs described
above.  All vibrational states up to $v=17$ ($X$) and $v=9$ ($a$),
rotational states up to $J=80$ and energies up to $D_0(X)$ were
considered. A dipole moment cutoff of $10^{-7}$~D was applied. The
zero point energy was calculated to be 1170.47 cm$^{-1}$ defined as the lowest state of the system ($J=1, v=0, N=0, e (-)$) above the minimum of the $X$ potential ($V_{\rm e} = 0$~\cm).  The final line list contains 2528 states and
65~055 transitions.  These levels are sufficient to represent the
ground state up to temperatures of about 4000~K but, at these
temperatures electronically excited states should also be occupied
giving rise to further transitions not considered here.  The line list
consists of the electric dipole transitions only. Moreover, the \X\
and \astate\ states are fully uncoupled and therefore the line list
does not include the weak, forbidden $a-X$ transitions observed in
\citet{96BeSeSh.PH}.

The results are provided in the standard ExoMol format \citep{jt631} as states and transitions files,
see extracts given in Tables~\ref{t:states} and \ref{t:trans}, respectively. Since \duo\ works
in Hund's case (a) but PH is a Hund's case (b) molecule for the $X$ state, the $(\pm,\Lambda,\Sigma,\Omega)$ quantum numbers were converted to $N,e/f$. For the \astate\ state,  $N=J$, $\Sigma=0$, $\Lambda=\Omega=2$. Here $N$ is the rotational quantum number defined as a projection of $\vec{N} = \vec{J}-\vec{S}$, where $\vec{J}$ and $\vec{S}$ are the total and spin angular momenta, respectively. The states file also gives Land\'{e} $g$-factors for the various states
\citep{jt656} which can be used to model the behaviour of these states in
a weak magnetic field.

\begin{table}\centering
{\tt
\caption{Sample extract of the states file for $^{31}$PH.}
\begin{tabular}{rrrrrcclrr}
\hline\hline
           $i$          &   \multicolumn{1}{c}{$\tilde{E}$}    &      $g_{i}$   &   $J$ & \multicolumn{1}{c}{$g$} &	$\pm$ &$e/f$    &State	&	$v$	& $N$  \\
\hline
    73  &      16.481110  &   20  &      2  &     1.00004  &   +     &     e     &    X3Sigma-    &      0   &       1   \\
    74  &     100.157012  &   20  &      2  &    -0.66632  &   +     &     e     &    X3Sigma-    &      0   &       3   \\
    75  &    2292.186846  &   20  &      2  &     0.99997  &   +     &     e     &    X3Sigma-    &      1   &       1   \\
    76  &    2373.329250  &   20  &      2  &    -0.66625  &   +     &     e     &    X3Sigma-    &      1   &       3   \\
    77  &    4480.481941  &   20  &      2  &     0.99989  &   +     &     e     &    X3Sigma-    &      2   &       1   \\
    78  &    4559.093532  &   20  &      2  &    -0.66618  &   +     &     e     &    X3Sigma-    &      2   &       3   \\
    79  &    6581.109731  &   20  &      2  &     0.99981  &   +     &     e     &    X3Sigma-    &      3   &       1   \\
    80  &    6657.178701  &   20  &      2  &    -0.66610  &   +     &     e     &    X3Sigma-    &      3   &       3   \\
    81  &    7573.601437  &   20  &      2  &     0.66667  &   +     &     e     &    a1Delta     &      0   &       2   \\
    82  &    8593.416817  &   20  &      2  &     0.99974  &   +     &     e     &    X3Sigma-    &      4   &       1   \\
    83  &    8666.923664  &   20  &      2  &    -0.66602  &   +     &     e     &    X3Sigma-    &      4   &       3   \\
    84  &    9899.143295  &   20  &      2  &     0.66667  &   +     &     e     &    a1Delta     &      1   &       2   \\
    85  &   10516.156390  &   20  &      2  &     0.99966  &   +     &     e     &    X3Sigma-    &      5   &       1   \\
    86  &   10587.067513  &   20  &      2  &    -0.66594  &   +     &     e     &    X3Sigma-    &      5   &       3   \\
\hline
\end{tabular}\label{t:states}

\mbox{}
}
{\flushleft
$i$:   State counting number.     \\
$\tilde{E}$: State energy in \cm. \\
$g_i$:  Total statistical weight, equal to ${g_{\rm ns}(2J + 1)}$.     \\
$J$: Total angular momentum.\\
$g$: Land\'{e} $g$-factors. \\
$e/f$:   Rotationless parity. \\
$\pm$:   Total parity. \\
State: Electronic state.\\
$v$:   State vibrational quantum number. \\
$N$: Rotational quantum number; $\vec{N} = \vec{J}-\vec{S}$ ($\vec{S}$ is spin).\\
}
\end{table}

\begin{table}\centering
{
\tt
\caption{Sample extract of the transitions file  for $^{31}$PH.}
\begin{tabular}{rrrr}
\hline\hline
    $f$          &   $i$    &      A$_{fi}$   &  \multicolumn{1}{c}{$\tilde{\nu}_{fi}$}  \\
\hline
          100     &      134 & 2.1325E-06     &      31.942920  \\
          162     &      135 & 3.0255E-05     &      31.959112  \\
          102     &       60 & 2.8977E-05     &      32.215418  \\
          256     &      285 & 3.2537E-06     &      32.223799  \\
           44     &        4 & 1.7832E-04     &      32.237083  \\
          174     &       75 & 6.7300E-04     &      32.439634  \\
          121     &       79 & 8.4497E-05     &      32.564436  \\
          119     &       20 & 5.1188E-04     &      32.625072  \\
         2047     &     2069 & 1.1922E-11     &      32.678476  \\
\hline\hline
\end{tabular}\label{t:trans}
\mbox{}
}
{\flushleft
$f$=state number of final state\\
$i$=state number of initial state\\
$A_{fi}$=Einstein-A coefficient\\
$\tilde{\nu_{fi}}$=transition wavenumber\\
}
\end{table}

\subsection{Partition Function}

The partition function, $Q(T)$ was calculated by summing the energy
levels given by \duo\ for temperatures up to $T=4000$~K. ExoMol follows the  HITRAN convention \citep{jt692},
of explicitly including the full atomic nuclear spin in the molecular partition function via the nuclear spin statistical number $g_{\rm ns}$. Since both P and H have nuclear spins $1/2$, $g_{\rm ns} = 4$.

The partition function at a range of temperatures is catalogued in
Table~\ref{t:Q}. It was compared with  sources where the partition
function $Q(T)$ was deduced from polynomial approximations \citep{84SaTaxx.partfunc,81Irwin.partfunc,16BaCoxx.partfunc}.
The cited  partition functions were all multiplied by 4 to match with the HITRAN
convention adopted. It can be seen that all the sources  approximately agree
other between 1000--4000~K. Below 1000~K, polynomial representations of $Q(T)$
used by \citet{84SaTaxx.partfunc} and \citet{81Irwin.partfunc} are not valid; our results
are much closer to the modern values of \citet{16BaCoxx.partfunc} albeit slightly higher probably
because of our full treatment of electron spin effects. Above 4000 K our values for $Q(T)$
are lower than those of  \citet{84SaTaxx.partfunc} and \citet{16BaCoxx.partfunc}; these works
include the contribution from electronically excited states which we neglect.


\begin{table}\centering
\caption{Comparison with calculated partition functions, $Q(T)$, of
 \citet{84SaTaxx.partfunc}, \citet{81Irwin.partfunc} and \citet{16BaCoxx.partfunc}.}
\label{t:Q}
\begin{tabular}{rrrrr}
\hline\hline
 $T$ / K           &   This work    &      Sauval \&\ Tatum   &    Irwin  & Barklem \&\ Collet\\
\hline
     100  &       103.24  &     112.20  &      12.03  &        103.33  \\
     200  &       202.58  &     222.11  &     125.10  &        202.68  \\
     300  &       302.14  &     320.89  &     261.37  &        302.25  \\
     400  &       401.99  &     416.37  &     383.16  &                \\
     500  &       502.54  &     512.03  &     494.35  &        502.64  \\
     600  &       604.56  &     609.72  &     600.96  &                \\
     700  &       708.99  &     710.52  &     706.92  &        709.08  \\
     800  &       816.74  &     815.16  &     814.60  &                \\
     900  &       928.57  &     924.17  &     925.49  &                \\
    1000  &      1045.12  &    1037.97  &    1040.52  &       1045.20  \\
    1500  &      1713.31  &    1689.58  &    1694.33  &       1713.48  \\
    2000  &      2549.16  &    2501.19  &    2499.05  &       2550.26  \\
    2500  &      3580.45  &    3500.15  &    3467.86  &                \\
    3000  &      4834.19  &    4713.74  &    4609.02  &       4847.12  \\
    4000  &      8112.16  &    7901.37  &    7436.84  &       8180.60  \\
\hline
\end{tabular}
\end{table}

\subsection{Experimental spectra}

\begin{figure}
    \centering
    \label{f:Bernath}
\includegraphics[width=0.45\textwidth]{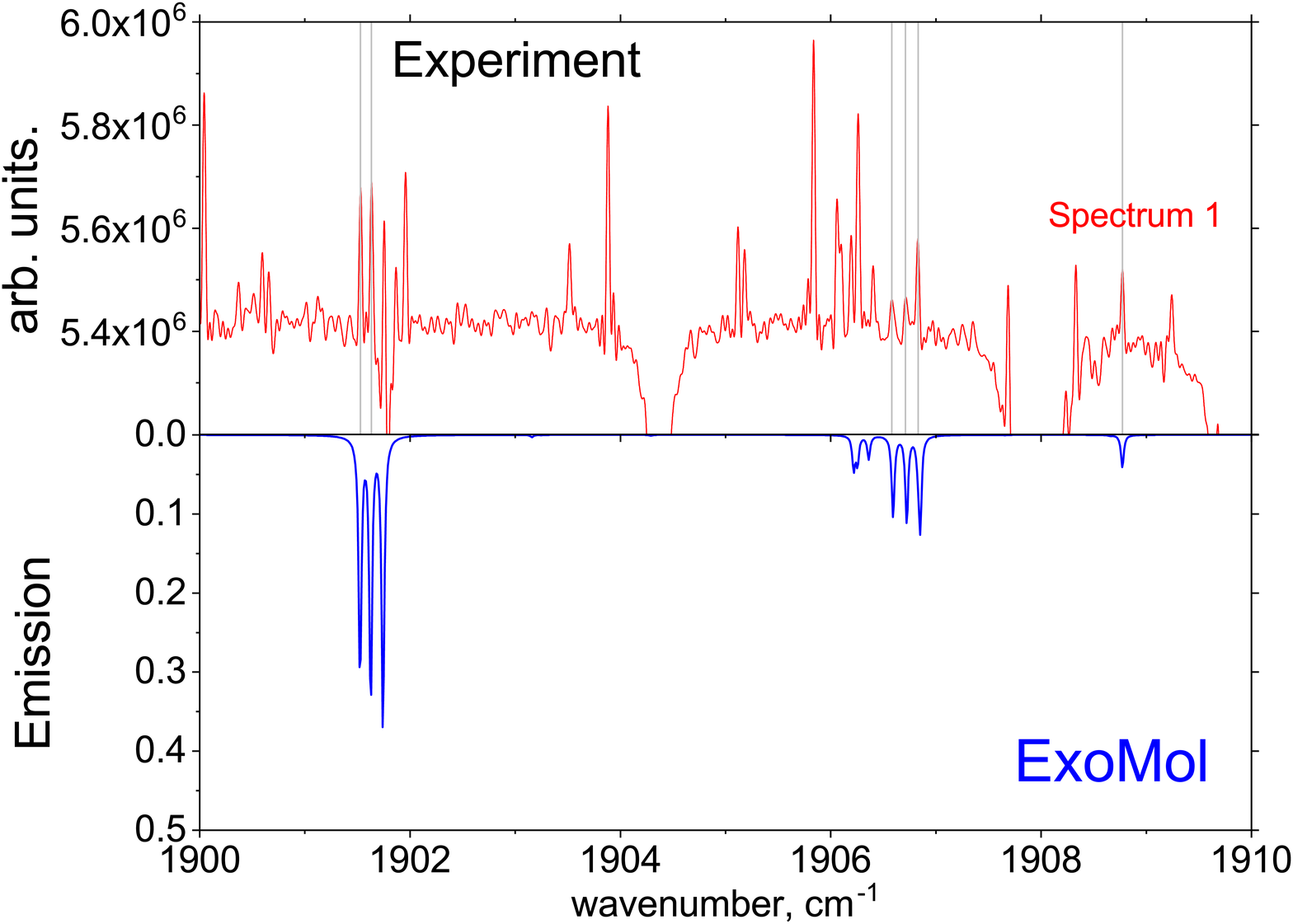}
\includegraphics[width=0.45\textwidth]{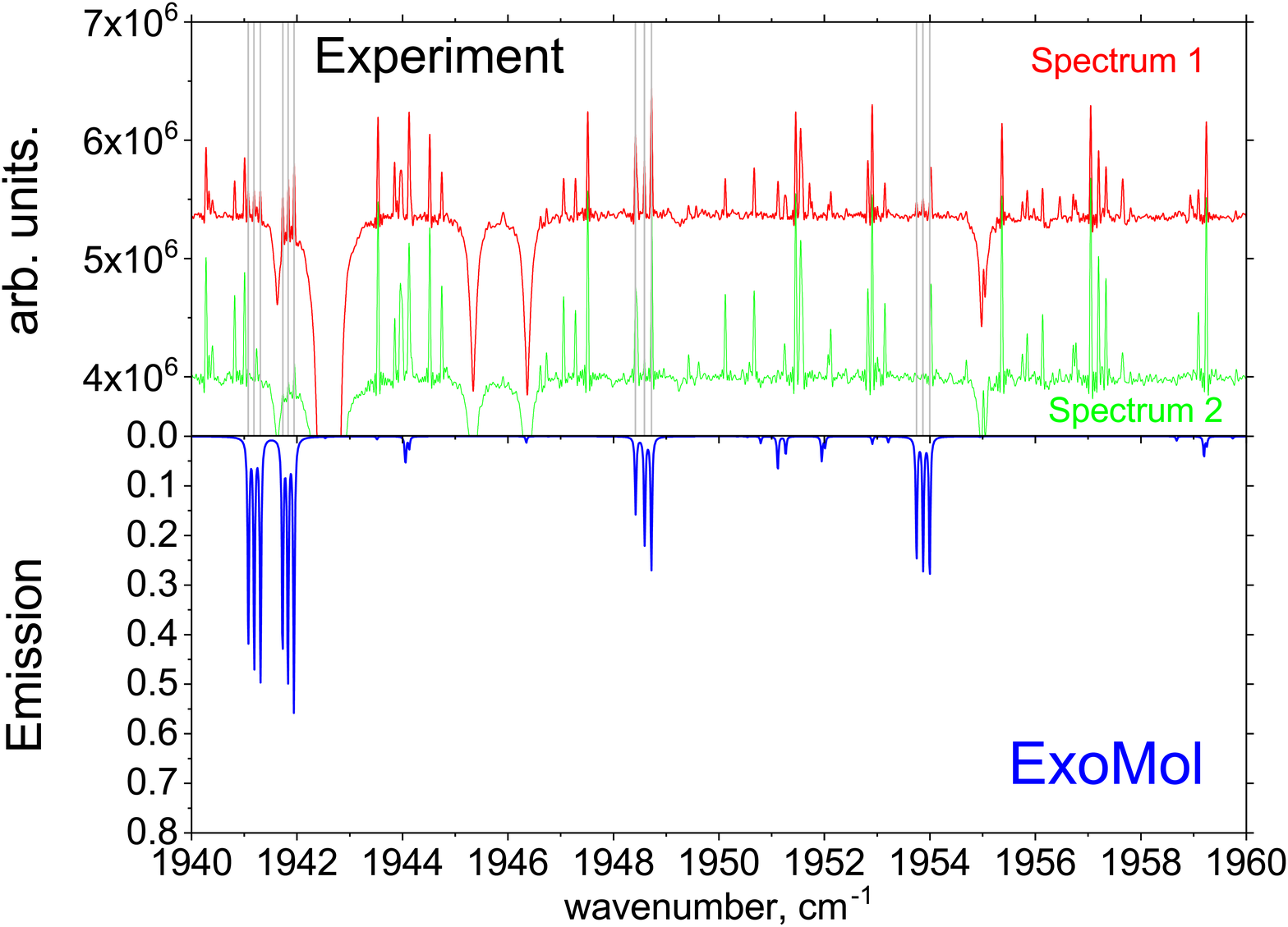}
\includegraphics[width=0.45\textwidth]{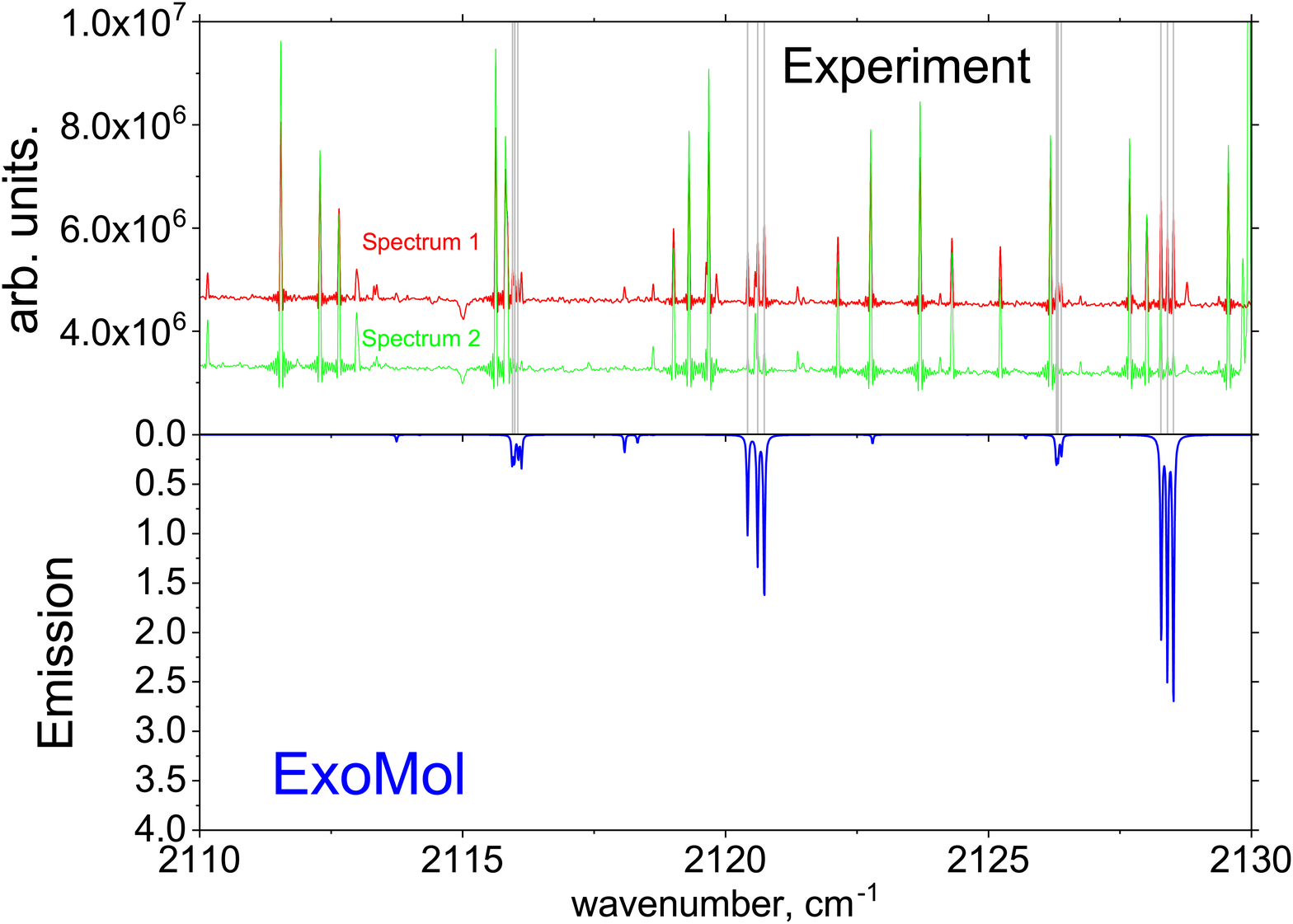}
\includegraphics[width=0.45\textwidth]{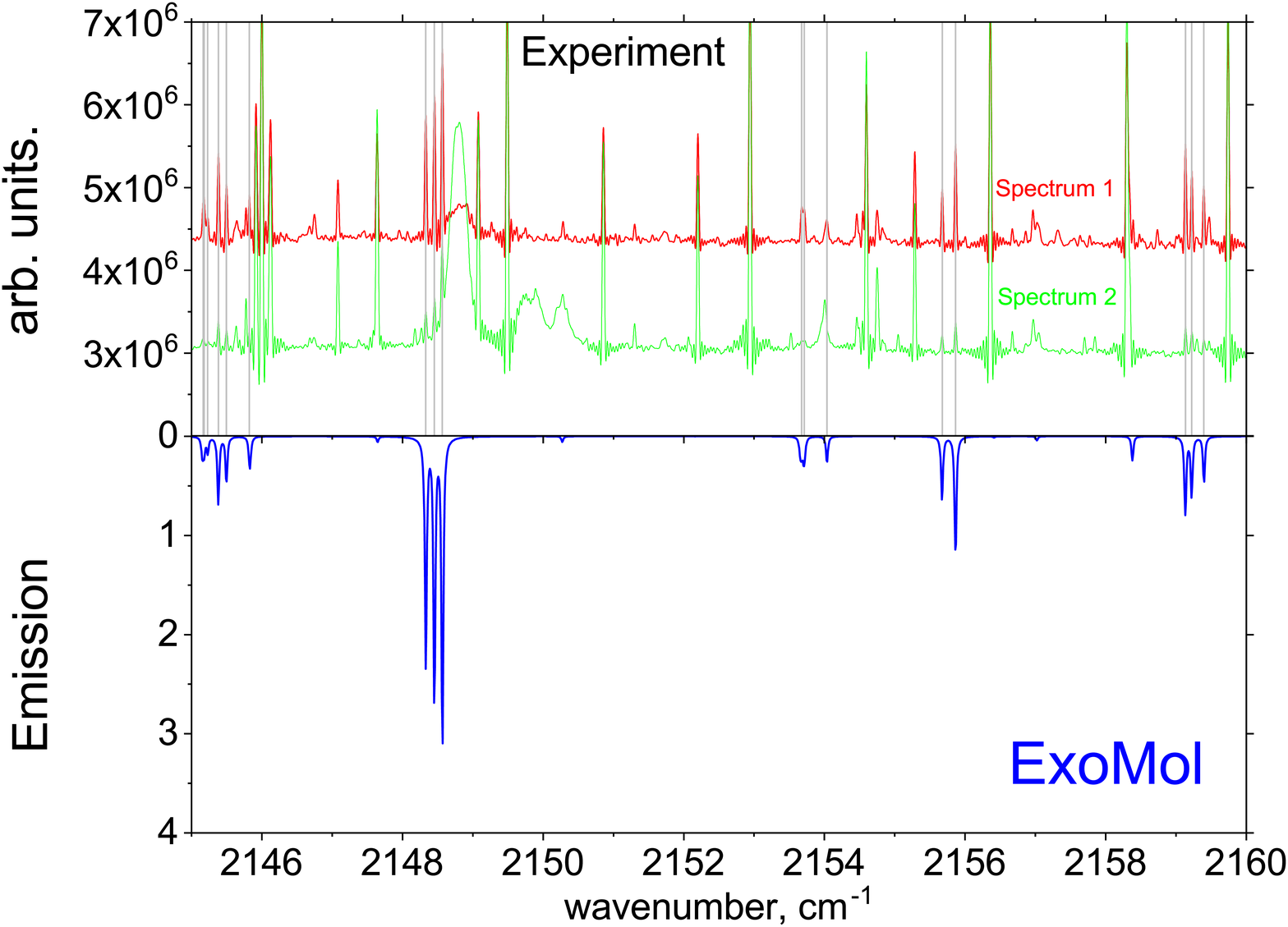}
\includegraphics[width=0.45\textwidth]{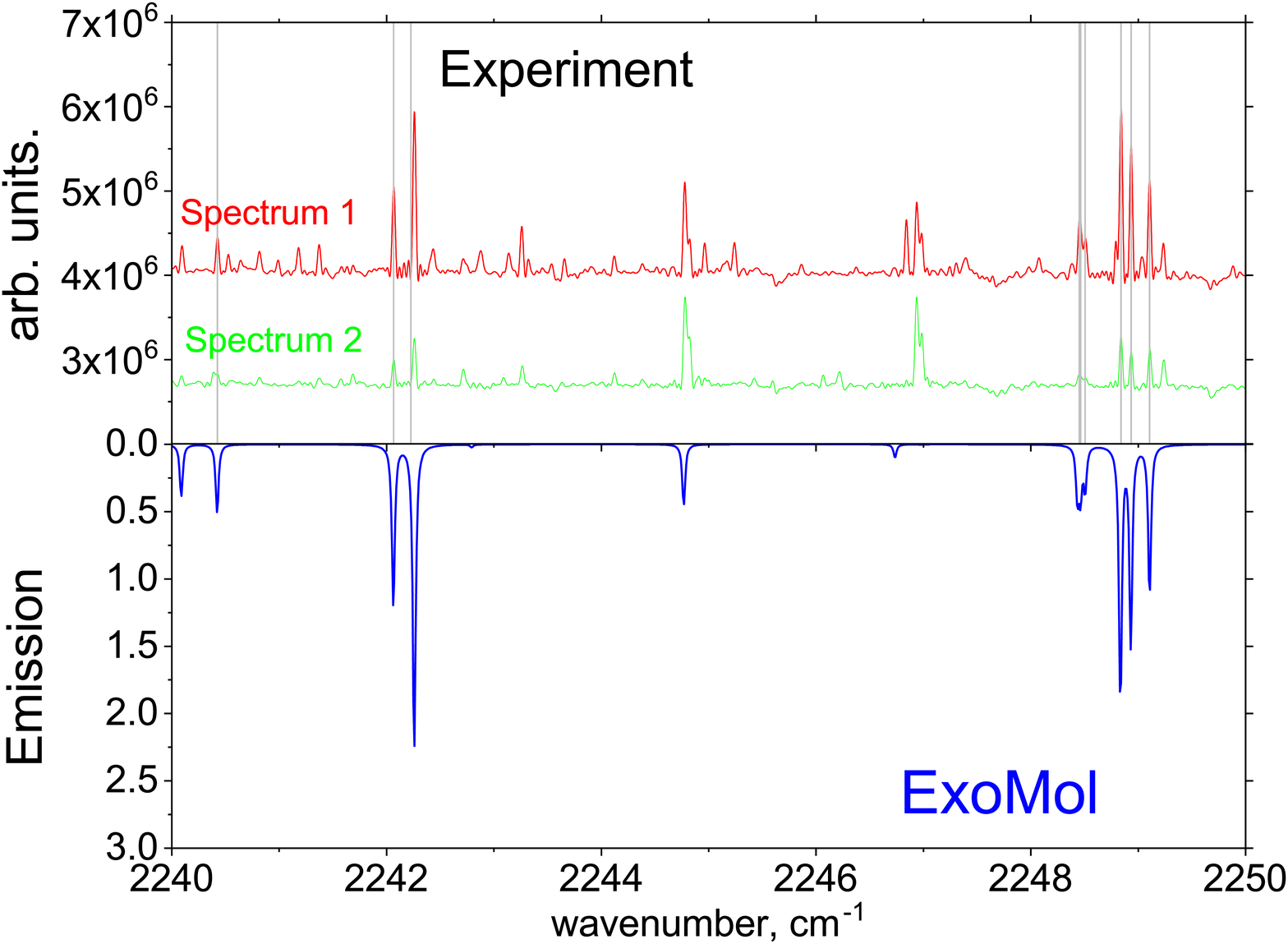}
\includegraphics[width=0.45\textwidth]{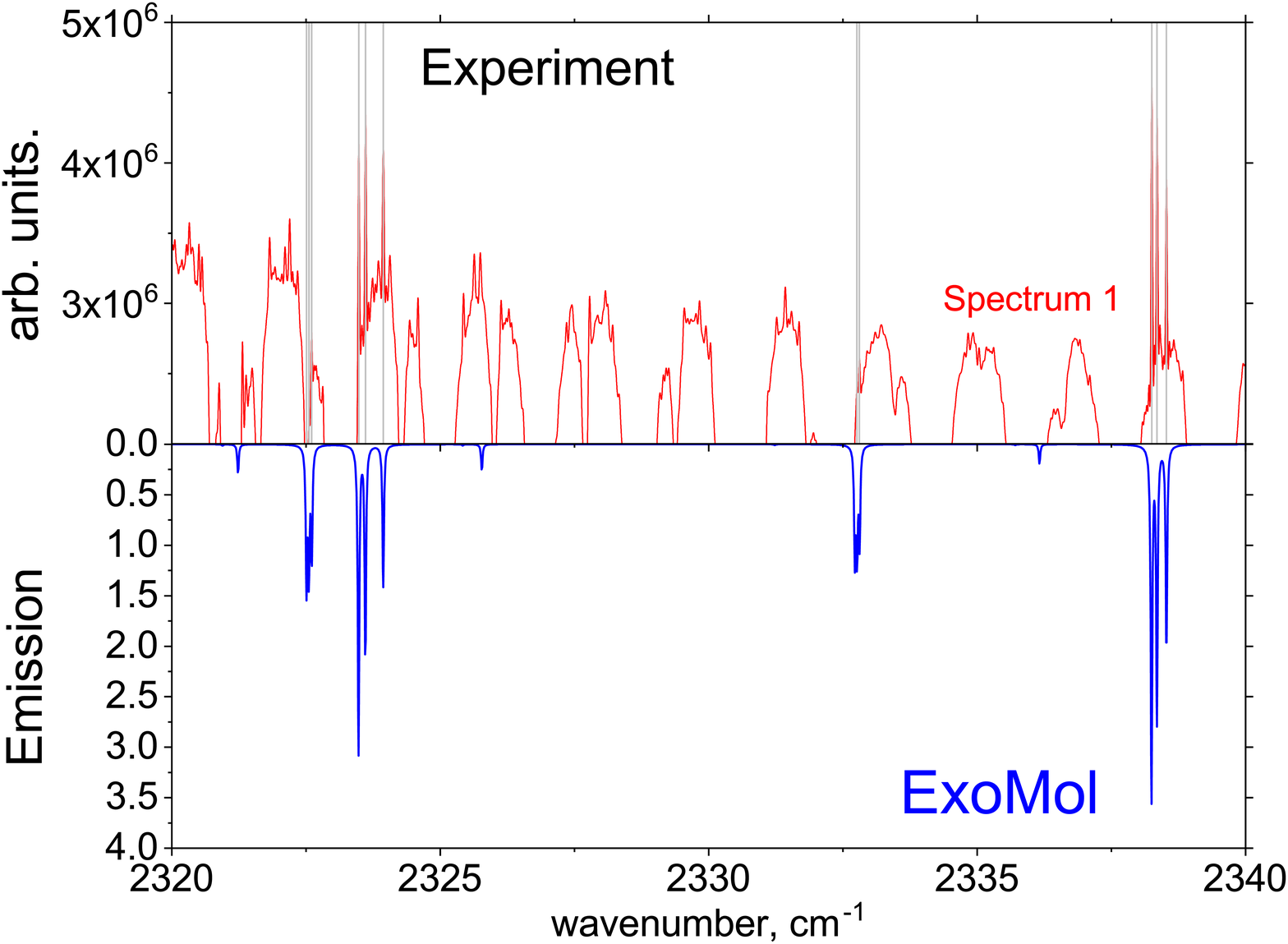}
    \caption{Experimental infrared spectra (see text for details)  of the \X\ state of PH
    compared to the ExoMol emission (photons/s) at $T_{\rm rot}=800$~K and $T_{\rm vib}=2300$~K using a Voigt line profile with $\gamma=0.15$~\cm. Vertical grey lines indicate experimental line positions from \citet{87RaBe.PH}.   }
\end{figure}

Two vibration-rotation emission spectra were recorded with the Fourier
transform spectrometer at the National Solar Observatory at Kitt Peak
in Arizona. The first spectrum (Spectrum 1 shown as the upper red
trace in Fig.~\ref{f:Bernath}) \citep{87RaBe.PH} was recorded with an
electrodeless quartz discharge tube excited with a 2450-MHz microwave
oscillator. A mixture of 0.45 Torr of hydrogen and 0.04 Torr of white
phosphorus vapor flowed through the cell. The spectral resolution was
0.02 \cm\ and covered the 1800--9000 \cm\ region. The second spectrum
(Spectrum 2 shown as the lower green trace in Fig.~\ref{f:Bernath}) differed only in
the gas mixture used: 2.75 Torr of helium, 0.04 Torr of white
phosphorus and 0.03 Torr of methane. Both spectra contained many
molecules including CO, CH, PH, CP, P$_2$, ArH, CN and C$_2$. The
strongest interfering molecule overlapping PH is CO with its $\Delta
v$=1 emission lines. The CO lines were used for wavenumber
calibration. The vibration-rotation emission lines of PH are stronger
in the first spectrum but the number of interfering lines is somewhat
reduced in the second spectrum. Norton-Beer strong apodization was
used because the lines were not resolved at 0.02 \cm\ resolution and
the lines still have residual ``ringing'' from the instrument line
shape function.

\subsection{Theoretical spectra}

\begin{figure}
    \centering
    \label{f:overview}
   \includegraphics[width=0.75\textwidth]{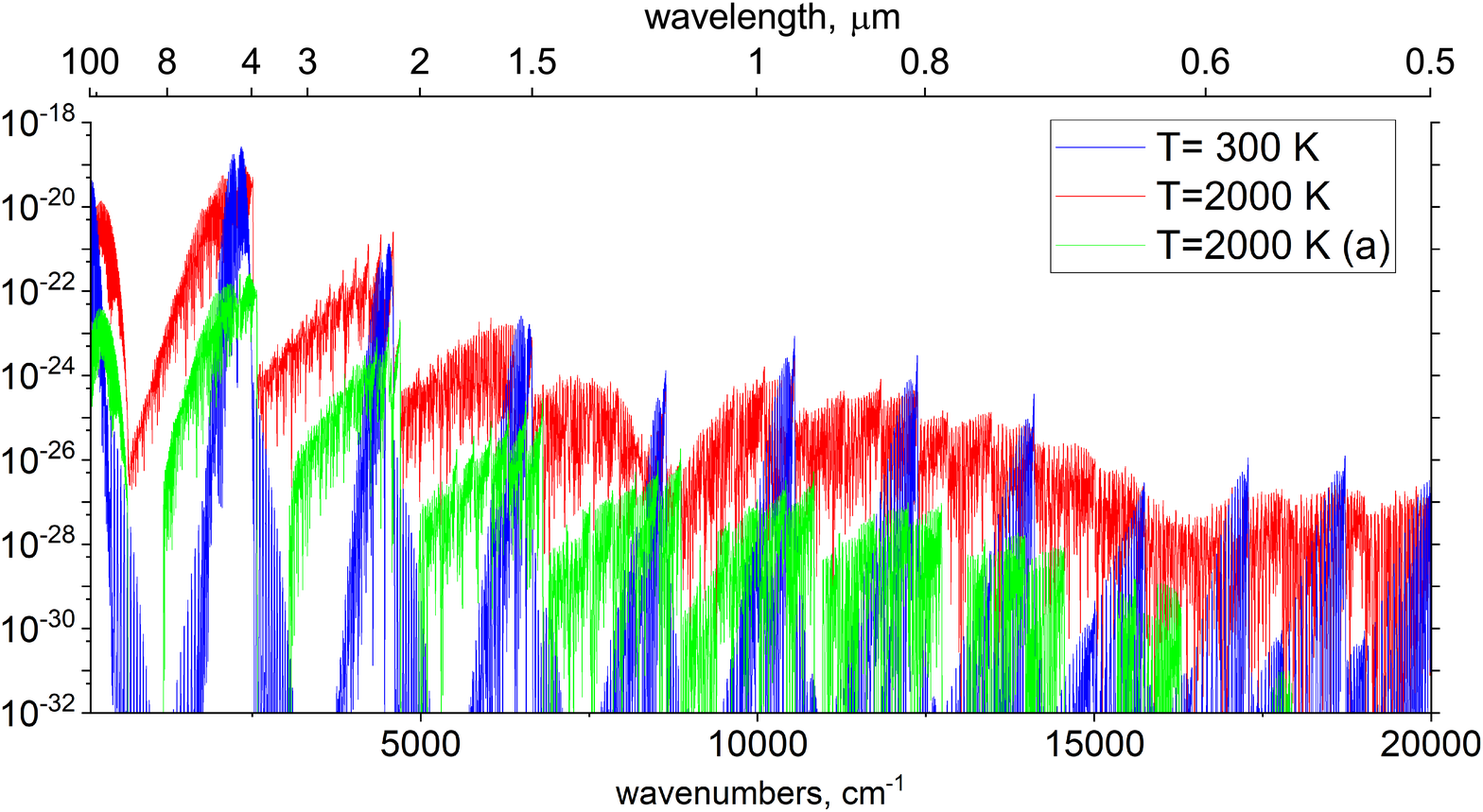}
    \caption{Absorption spectrum of PH at different temperatures computed using ExoMol line list and a Gaussian line profile of 2 \cm. The contribution from the $a$-band at $T=2000$~K is shown separately.}
\end{figure}

Figure \ref{f:overview} gives an overview of  full PH spectrum in absorption at two temperatures, 300~K and 2000~K.
At the lower temperature a clear progression of vibrational bands can be seen which is substantially washed out
at the higher temperature due to the presence of many more weak lines in the spectrum.

\begin{figure}
    \centering
   \includegraphics[width=0.8\textwidth]{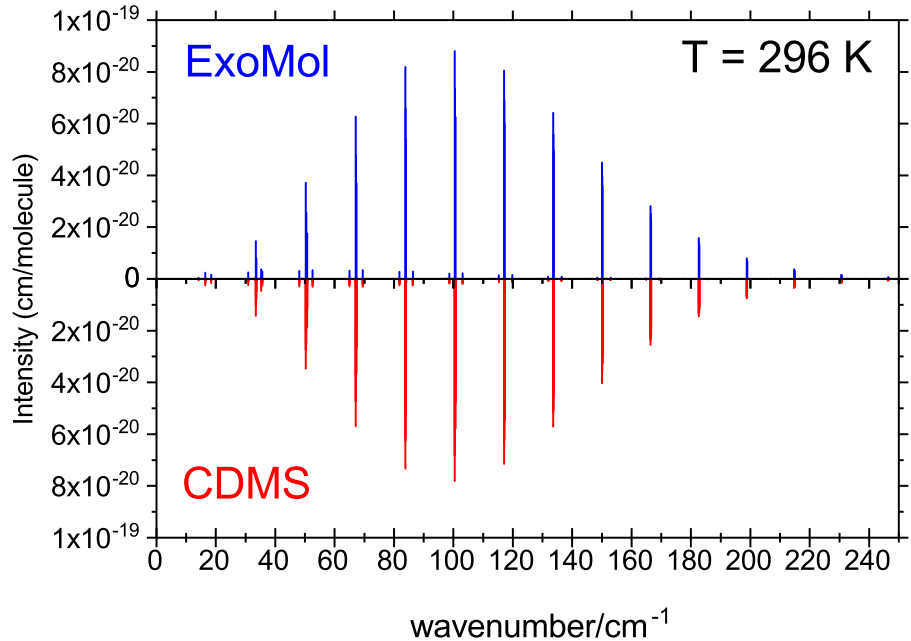}
    \label{f:CDMS}
    \caption{Pure rotational band within $v=0$ for the rovibrational absorption stick spectra of the \X\ state of PH at 296~K.}
\end{figure}

Figure \ref{f:CDMS} shows the microwave spectrum of the ground state of PH at
$T$=296~K.  It is compared with the CDMS  (Cologne Database for Molecular
Spectroscopy) spectrum \citep{cdms} ($v=0$), which we have averaged over the hyperfine components. The dipole moment used by CDMS  is 0.396~D at $r_{\rm e}$ (M\"{u}ller, unpublished work), whereas the
dipole moment from this work is 0.477~D at equilibrium or, more importantly, our
$v=0$ state vibrationally-averaged dipole is 0.4499~D. This
disparity can be seen in the graph, with our lines being more intense
due to the larger dipole moment. Partition function values from CDMS agree with that
from this work. At 300~K, the CDMS value was 302.12, while our value is 302.14.  The line strength depends on the dipole
moment squared. With this in mind, our intensities are about 1.27 times higher than those given by CDMS.
We suggest that CDMS may wish to re-scale their intensity values to ours. Conversely, the CDMS frequencies
rely directly on data taken in the same frequency region and include the hyperfine components and must be considered more accurate than ours in this region; they should be used for any attempts to detect PH in the interstellar medium.

Figure \ref{f:Bernath} compares the infrared ro-vibrational experimental
spectra from \citet{87RaBe.PH} with calculations from this work. We assumed the non-LTE model based on two temperatures, rotational $T_{\rm rot}$ and vibrational $T_{\rm vib}$, as implemented in \Duo. The temperatures were adjusted to match the experimental emission spectra of PH, with the final values of $800$~K and 2300~K, respectively. The grey vertical line indicate the experimental line positions from \citet{87RaBe.PH}.
The good agreement with the observed spectrum of \citet{87RaBe.PH} is found across the entire region and confirms the accuracy of our calculations.

\section{Conclusion}

A comprehensive line list for the ground ($X$) and first excited ($a$) electronic states of $^{31}$PH, known as LaTY,
 is presented. It is based on an accurate PECs, BOB, SS and SR curves obtained by fitting to a set of experimental transition line frequencies and extrapolating to higher ro-vibrational levels and \ai\ dipole moment curves.
Future work can include analysis and investigations of higher electronic states
of PH with the aim of creating a line list appropriate at shorter wavelengths. Previous  studies on the low-lying
$b^1{}\Sigma^+$ of PH \citep{78StLeMa.PH,78NgStLe.PH,99StLe.PH} as well as on the strongly dipole-allowed $A$~$^3\Pi$ -- \X\ system  \citep{02FiChMo.PH,03FiChWe.PH,14GaGa.PH}  in the near-ultraviolet provide a good starting point for future analysis of the PH molecule.

The line list can be downloaded from the CDS, via\\
ftp://cdsarc.u-strasbg.fr/pub/cats/J/MNRAS/, or
http://cdsarc.u-strasbg.fr/viz-bin/qcat?J/MNRAS/, or from
www.exomol.com.

\section*{Acknowledgements}

This work was supported by the UK Science and Technology Research
Council (STFC) No. ST/R000476/1 and the COST action MOLIM No. CM1405.
This work made extensive use of UCL's Legion high performance
computing facility  along with the STFC DiRAC HPC facility
supported by BIS National E-infrastructure capital grant ST/J005673/1
and STFC grants ST/H008586/1 and ST/K00333X/1. Some support was provided by the NASA Laboratory Astrophysics program.

\bibliographystyle{mn2e}

\label{lastpage}

\end{document}